# The Dynamic Spectrum Aggregation Strategy for Cognitive Networks Based on Markov Model

Yifei Wei, Qiao Li, Xia Gong, Da Guo and Yong Zhang
School of Electronic Engineering, Beijing University of Posts and Telecommunications, Beijing 100876, China
Email: weiyifei@bupt.edu.cn, liqiao1989@bupt.edu.cn, yongzhang@bupt.edu.cn

*Abstract*—**In order to meet the constantly increasing demand by mobile terminals for higher data rates with limited wireless spectrum resource, cognitive radio and spectrum aggregation technologies have attracted much attention due to its capacity in improving spectrum efficiency. Combing cognitive relay and spectrum aggregation technologies, in this paper, we propose a dynamic spectrum aggregation strategy based on the Markov Prediction of the state of spectrum for the cooperatively relay networks on a multi-user and multi-relay scenario aiming at ensuring the user channel capacity and maximizing the network throughput. The spectrum aggregation strategy is executed through two steps. First, predict the state of spectrum through Markov prediction. Based on the prediction results of state of spectrum, a spectrum aggregation strategy is proposed. Simulation results show that the spectrum prediction process can observably lower the outage rate, and the spectrum aggregation strategy can greatly improve the network throughput.**

## I. INTRODUCTION

In recent years, the development of wireless communication technologies and the high data rate services has consumed almost all of the available spectrums, so the spectrum resource has become a very scarce radio resource. And according to the existing spectrum allocation policy, most of the spectrums are assigned to specific authorized users or services, so it becomes more difficult to find a bandwidth to meet new demands within the available spectral range [1]. Meanwhile, the fixed spectrum allocation strategy [2] is an inefficient use of spectrum both in time and space. Through the spectrum sensing technology of cognitive radio, users who have transmission needs are enabled to access to the ideal spectrum dynamically, thus improve the spectrum utilization efficiency [3].

However, limited by the communication development history and the shortage of wireless spectrum resource, most of the available idle spectrums are non-continuous, and the single spectrum is difficult to meet the bandwidth requirement of LTE-Advanced [4]. Therefore, in order to meet the demands of diverse services and transmission rate, LTE-Advanced put forward the concept of spectrum aggregation [5]. The main idea of spectrum aggregation technology is to obtain a wider spectrum by merging a number of continuous or discrete spectrums together [6][7][8]. The application of this technology can not only meet the LTE-Advanced requirements, but also can improve the utilization rate of spectrum fragmentation [9]. On the other hand, in the actual utilization of spectrum, there are a lot of unused discrete spectrums among the authorized spectrum. But these idle spectrums can't be obtained by secondary users [10], thus caused the waste of spectrum on the condition of the shortage of spectrum resource. So making research of spectrum aggregation technology has a very important practical significance in satisfying the transmission requirement of secondary users [11] [12].

In the actual networks, while the number of users who have a large bandwidth demands is growing, the traditional continuous spectrum allocation strategy can't make full use of the existing discrete spectrum resources in the networks, and the continuous spectrum resources will be split, thus leading to an increased number of spectrum fragmentation. Therefore, when the spectrum fragmentations are increasing, we can make use of these spectrum fragmentations by using the spectrum aggregation technology, and achieve the goal of improving the utilization of spectrums and the data transmission rate, especially in the cell edge [13].

### A. *Related Work*

According to the research, we found out that the existing spectrum aggregation technology in cognitive cooperation networks can be roughly divided into two categories: the spectrum aggregation in single-relay networks and that in multi-relay cooperative networks [14].

#### *1) Spectrum aggregation technology in single-relay networks[15]*

In a single-relay network, a source node can send data to the destination node directly or through relays. Because of the transmission distance, the communication between two far nodes will be affected from the channel fading and lead to transmission failure. At this time, through the relay node, the data can be successfully transmitted to the destination node, so the relay forwarding have expanded the communication distance and have improved the network reliability. In a single-relay network scenario shown in Figure 1, users can choose the direct transmission mode or cooperative relay transmission mode with a relay node adaptively.

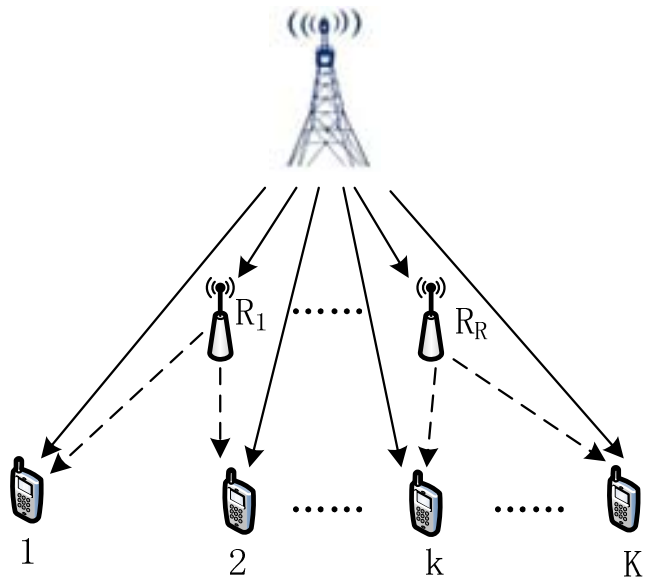

Figure 1. Single-user transmit data through single-relay

The basic idea of spectrum aggregation algorithm in single-relay network is that shows below. At first, choose a relatively good transmission mode according to the network performance. If there is a need to use a relay to forward data, select an optimal relay. And then search the available bandwidth from the low frequency to the high frequency of the whole system. Finally aggregate the discrete spectrums on the conditions of the spectrum aggregation span. The spectrum aggregation scheme has improved the transmission rate, the spectrum utilization, the throughput and performance of the system.

*2) Spectrum aggregation technology in multi-relay networks[16][17]*

Combined the multi-relay cooperative technology with the spectrum aggregation, we consider a network scenario as shown in Figure 2. In the network, a secondary user S (source node) has a need to send data to a secondary user D (destination node), and because of the transmission distance or others, the data has to forward by relay nodes.

The basic idea of spectrum aggregation in multi-relay cooperative networks [18] can be described below. At first, the source node and relay nodes get the spectrum state message by spectrum sensing technology. And then as for every common ideal spectrum of the source node and relay nodes, select an optimal relay based on the system performance. Finally aggregate all the spectrums that selected a same relay node.

## B. Contributions and Orgnization

However, in the above spectrum aggregation schemes of single-relay and multi-relay networks, they both only considered the situation of single-user, and ignored the actual transmission needs of multi-user in a real network. And there is no consideration and analysis of the overall network performance. Therefore, in this paper, we discuss the spectrum aggregation scheme of multi-user cooperative relay network aimed to improve the spectrum efficiency and data transmission rate, and analyze the impact on the overall network performance, such as network throughput and channel capacity. Based on Markov prediction we design a spectrum prediction model in multi-user cooperative relay networks, then the spectrum aggregation strategy aiming at improving the spectral efficiency is proposed based on the first step operation of spectrum prediction.

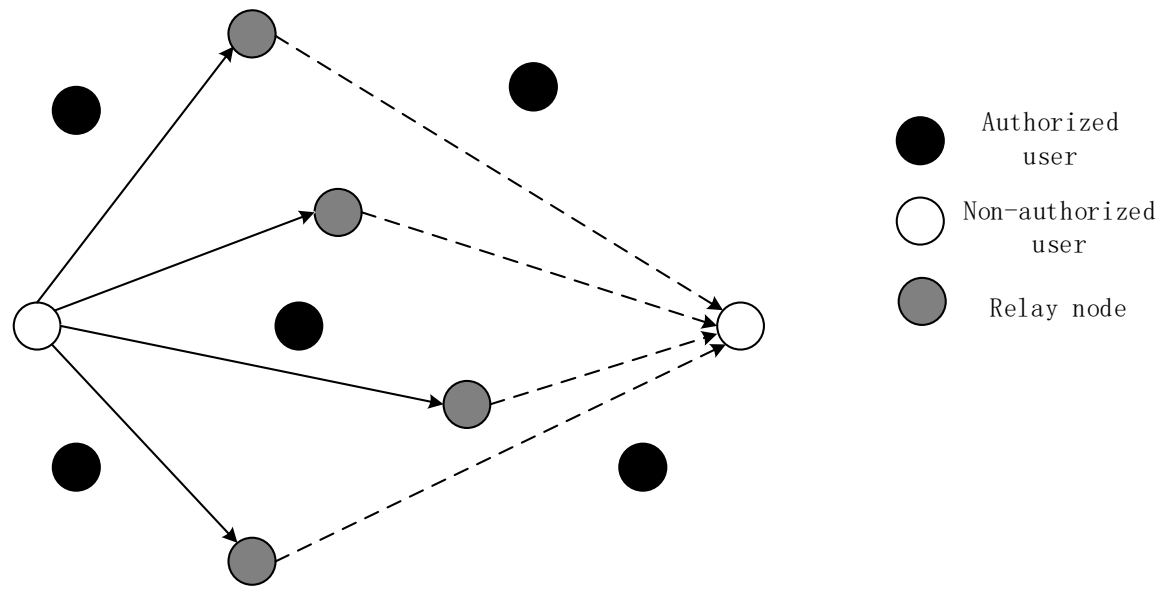

Figure 2. Single-user transmit data through multi-relay

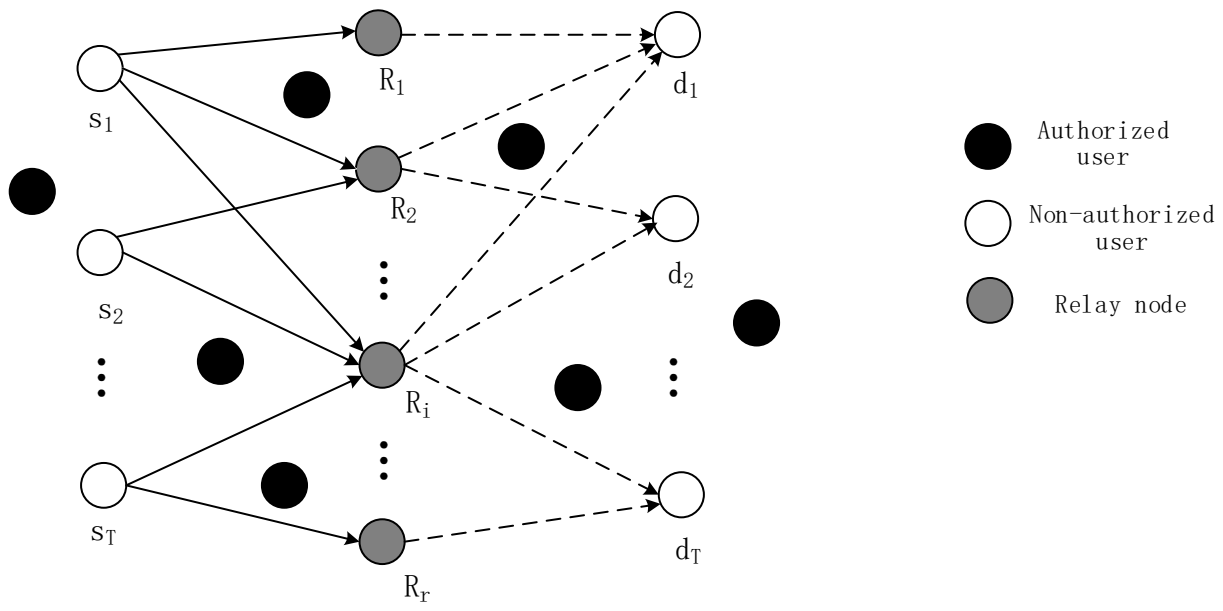

Figure 3. Multi-user transmit data through multi-relay

This paper is organized as follows. System model is provided in section II. Section III emphatically expounds the spectrum aggregation algorithm. The simulation results and related discussion is provided in section IV. Finally, we conclude this paper in section V.

# II. System Model

As is shown in Figure 3, we consider a two-hop network with multiple users who have transmission need simultaneously, the source nodes $S=\{s_1,s_2,...,s_T\}$ send data to the corresponding destination nodes $D=\{d_1,d_2,...,d_T\}$ . In this network, due to the limitation of distance, the source nodes have to send data to the destination node through a relay node.

Assume that the total system bandwidth is $B$ , and the total spectrum is divided into $N$ sections, so the spectrum bandwidth of each segment is $b_n = B/N$ . When the authorized users have transmission need, the corresponding licensed spectrum will be occupied.

As we can see from Figure 3, the relay nodes is $\{r_1,r_2,...,r_i,...,r_r,\}$ , and the relay nodes in the coverage of the source node $s_t$ and the destination node $d_t$ is $R_{S_t}$ .In the relay cooperative communication, we take DF mode into consideration, so the signal received by the destination nodes d ( $d=d_1,d_2,...,d_T$ )is as follows:

$$y_{d,n} = h_{(r_i,d),n}\sqrt{P_{(r_i,d),n}}\,x_{d,n} + n_{d,n} \tag{1}$$

where $x_{d,n}$ is the signal sent to the destination node d using the band $n$ , $y_{d,n}$ is the signal received by the destination node d using the band $n$ , $n_{d,n}$ is the noise signal, $h_{(r_i,d),n}$ and $P_{(r_i,d),n}$ represent the channel gain and transmission power through the relay $r_i$ to the destination node d using band $n$ .

In this network scenario, two time slots are needed to complete a transmission process from a source node to the destination node. Spectrum sensing and transmission from the source nodes to the relay nodes can be complicated in the first time slot. In the second slot, relay nodes forward the data to the destination nodes.

At the beginning of the first slot, the source nodes and the relay nodes conduct spectrum sensing to get the available spectrum bands, and we can get the spectrum states of the source nodes and the relay nodes,

Source node:

$$G = \{g_n^s \mid g_n^s \in \{0,1\}, n = 1,2,...,N\} \qquad (2)$$

Relay node:

$$H_i = \{h_n^i \mid h_n^i \in \{0,1\}, n = 1,2,...,N\}, \forall i \in \{1,2,...,R_r\} \qquad (3)$$

As is shown in formula (2) and (3), if a source node or relay node perceive that the spectrum $n$ is idle, then let $g_n^s = 0$ or $h_n^i = 0$ , otherwise, let $g_n^s = 1$ or $h_n^i = 1$ . Therefore, the probability of 0 or 1 represent the probability of that the spectrum is free or occupied. As for the source nodes, the probability of spectrum idle or occupied can be represented respectively as:

$$p_0^s = \Pr[g_n^s = 0]$$
$$p_1^s = \Pr[g_n^s = 1] = 1 - p_0^s \qquad (4)$$

And for the relay nodes, we can also get the probability of spectrum idle or occupied,

$$p_0^i = \Pr[h_n^i = 0]$$
$$p_1^i = \Pr[h_n^i = 1] = 1 - p_0^i \qquad (5)$$

Suppose that in a same network, the probability of spectrum idle or busy of the source nodes and relay nodes are equal, and they can be expressed as $P_0$ and $P_1$ . $P_0$ and $P_1$ are system parameters, they can be affected by time, location and available spectrum and so on.

## III. THE PROPOSED ALGORITHM

### A. *Spectrum prediction model based on Markov model*

The spectrum aggregation algorithm described above is proceed in two time slots: the data transmission from source node to relay node is finished in the first time slot, and the data is retransmitted from relay node to destination node in the second time slot. The flowchart of communication process is illustrated in Figure 4. However, the wireless channel is time varying and the spectrum state is also dynamically changing so that the spectrum state of the second slot may be different from that been predicted in the first time slot. In order to solve the problems stated above, in this section we research the spectrum prediction model for the collaborative relay networks to lower the outage rate. Since Markov channel models have been widely accepted in the literature [19] [20] as an effective approach to characterize the correlation structure of the fading process, the Markov model is used here.

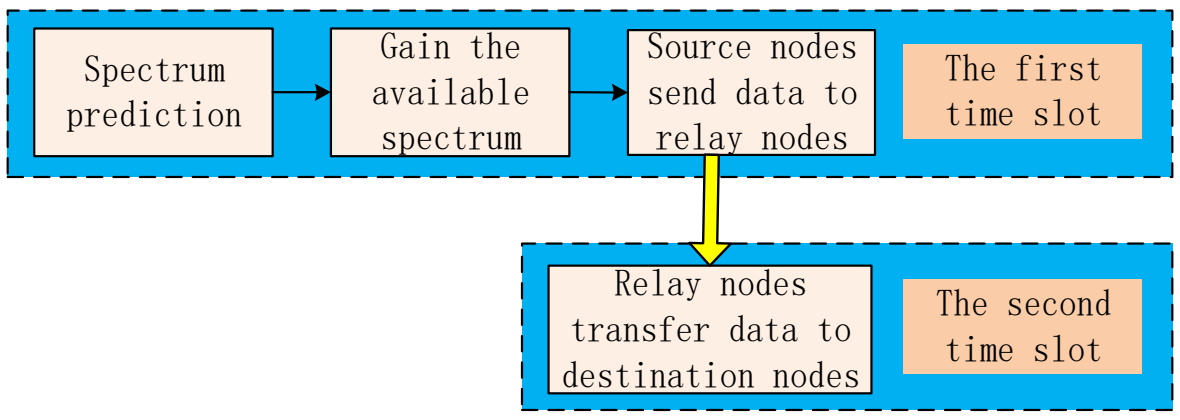

Figure 4. Flowchart of communication process in collaborative relay networks

#### *1) Markov Chain*

Markov chain is a Markov process with discrete time and state. As a stochastic process, $\{X_n, n \in T\}$ , the state space is I, and $T = \{0,1,2,...\}$ . If a stochastic process is a Markov process, for any integer n and state $i_0, i_1, ..., i_{n+1} \in I$ , the following equation is established:

$$P\{X_{n+1} = i_{n+1} \mid X_0 = i_0, X_1 = i_1, ..., X_n = i_n\} = P\{X_{n+1} = i_{n+1} \mid X_n = i_n\} \qquad (6)$$

The state at t=n+1is only decided by the state at time t=n. The state-transition matrix is described as follows,

$$P = \begin{pmatrix} p_{11} & p_{12} & \cdots & p_{1m} \\ p_{21} & p_{22} & \cdots & p_{2m} \\ \cdots & \cdots & \cdots & \cdots \end{pmatrix} \qquad (7)$$

Set the probability of state transmission from tme t at state $X_t = i$ to time t+1 at state $X_{t+1} = j$ as $p_{ij}$ . If the initial state is $P_0$ , after n steps of transmission, the system state is $P^{(n)}$ ,

$$P^{(n)} = P_0 P^n \qquad (8)$$

The basic idea of using the Markov chain: First, give a precise description of the state space of the system, then give the probability distribution of each state and the state transmission matrix. Based on the initial state probability and the probability distribution of the state transmission matrix, we can predict the state probability at any time, then we can compare the predicted state probability with state probability from the actual observation.

#### *2) Spectrum Predicton model*

In general, the state of spectrum in the collaborative relay networks is decided by the primary user who uses licensed spectrum. When the primary user uses the licensed spectrum, the state of this spectrum is busy, and this spectrum can't be used by the secondary users. While the licensed spectrum isn't be used by the primary, it means that this spectrum is idle and can be used by the secondary users. However, when the licensed is idle, these still be two state of this spectrum according to the channel state. When the channel state is good, this spectrum is good. A bad channel state will degrade the success rate of the data transmission, which means that the state of the spectrum is bad too. Through the analysis above, we divided the state of spectrum into three conditions which respectively is busy, good, and bad.

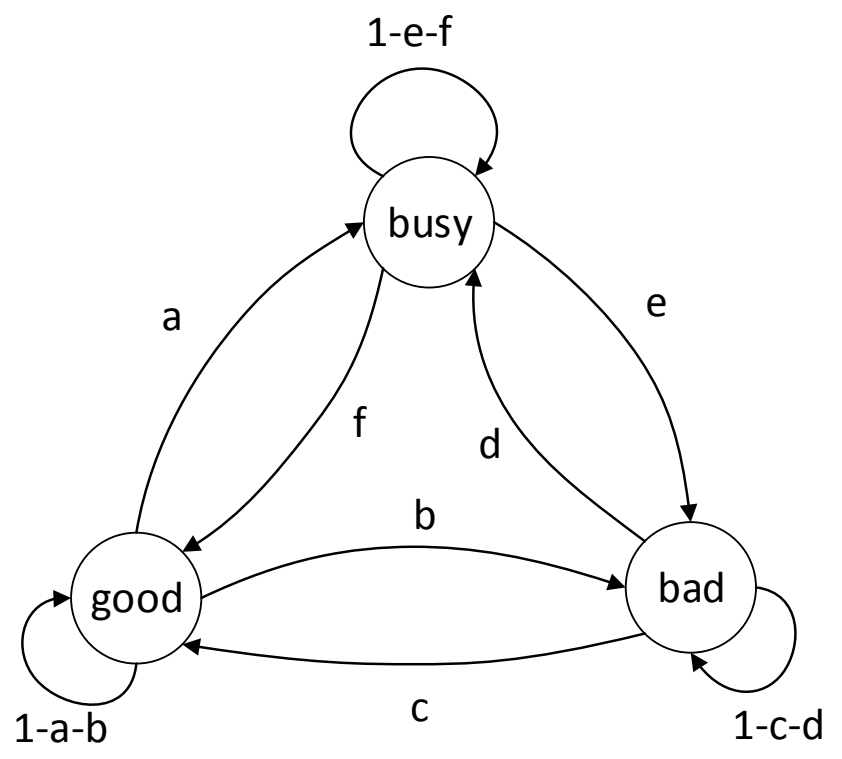

Figure 5. Markov model of spectrum state

We adopt the Markov model to predict the state of the spectrum and to get a relatively accurate state of the spectrum in next time slot, and according to this spectrum prediction, we make the spectrum aggregation strategy. In a multi-user cooperative relay network, we assume the state-transition matrix $P$ of the spectrum is written as follows when the system is in the steady state, as shown in Figure 5.

$$P = \begin{array}{c} \\ good \\ bad \\ busy \end{array} \begin{array}{c} \begin{array}{ccc} good & bad & busy \end{array} \\ \begin{pmatrix} 1-a-b & a & b \\ c & 1-c-d & d \\ f & e & 1-e-f \end{pmatrix} \end{array} \quad (9)$$

When the system is in the steady state, the stationary probability distribution is as follows,

$$\Pi = \{\pi_{good}, \pi_{bad}, \pi_{busy}\} \quad (10)$$

From equation (9) and (10), we get:

$$\begin{cases} \pi_{good} + \pi_{bad} + \pi_{busy} = 1 \\ (1-a-b)*\pi_{good} + c*\pi_{bad} + f*\pi_{busy} = \pi_{good} \\ a*\pi_{good} + (1-c-d)*\pi_{bad} + e*\pi_{busy} = \pi_{bad} \\ \text{b}*\pi_{good} + d*\pi_{bad} + (1-e-f)*\pi_{busy} = \pi_{busy} \end{cases} \quad (11)$$

We analysis the actual state of spectrum in N time slot, and let 1, 2, and 3 respectively represent the state of spectrum which is good, busy, and bad. Assuming N=20, we observe the state of the spectrum of one node: 00010210000112010001. According to the observed data, we get the First-order Markov state-transition matrix:

$$\begin{cases} p_{00} = \frac{7}{12} \\ p_{01} = \frac{4}{12} \\ p_{02} = \frac{1}{12} \end{cases} \quad \begin{cases} p_{10} = \frac{3}{5} \\ p_{11} = \frac{1}{5} \\ p_{12} = \frac{1}{5} \end{cases} \quad \begin{cases} p_{20} = \frac{1}{2} \\ p_{21} = \frac{1}{2} \\ p_{22} = 0 \end{cases} \quad (12)$$

We can predict the state of the spectrum in next slot using First-order Markov state-transition matrix. When the amount of the observe data is large enough, the state-transition matrix can accurately reflect the prediction result. However, the state of the spectrum in the network follows certain law decided by time. When we train the Markov model, the number of time slot N should not be too large.

After setting the number of time slot N, we model the Markov state-transition matrix N time slot before. Then we predict the state of the spectrum in the next slot through the Markov state-transition matrix. The detailed algorithm is described as follows,

1) Get the state of the spectrum St before N time slots according to the spectrum sensing process.
2) According to the Markov model, deduct the Markov state-transition matrix Pst.
3) Based on the current state of the spectrum and the Markov state-transition matrix Pst, get the state of spectrum at the next time slot.
4) According to the result of spectrum sensing process and the prediction result, let the spectrum which is idle at both two time slot as the public available spectrum.

### *B. Spectrum aggregation algorithm*

In this section, we proposed a spectrum aggregation algorithm in the collaborative relay network to maximize the network throughput [21]. Provided that the bandwidth of all spectrums are the same, namely $b_n = b$, and the procedure of the spectrum aggregation algorithm is described as follows,

*1) Choose the appropriate relay nodes $R_{s_t}$ for every user $s_t$.*

*for* $i = 1:r$
  *if* $r_i$ *only in range of* $s_t$ *and* $d_t$ *then*
    $R_{S_t} \leftarrow R_{S_t} + \{r_i\}$
  *else if* $r_i$ *can connect to several* $s_t(s_t = s_1, s_2, ..., s_T)$ *and* $d_t(d_t = d_1, d_2, ..., d_T)$ *then*
    $T_{r_i,d_t} = b\log_2(1 + \frac{P_{r_i,d_t} \times h_{r_i,d_t}}{\Gamma N_0 W})$
    $(d_t) \leftarrow \arg\max\{T_{r_i,d_t}\}$
    $R_{s_t} \leftarrow R_{s_t} + \{r_i\}$
  *else*
    *drop* $r_i$
  *end if*
*end*

where $T_{r_i,d_t}$ represents the throughput of the source node $s_t$ to the destination node $d_t$ through the relay $r_i$, $N_0W$ is the noise power of the system, $\Gamma$ is the redundancy associated with the physical code modulation, and it is related to bit error rate, namely $\Gamma = \frac{-1.5}{\log_2(5 \times BER)}$, $P_{r_i,d_t}$ and $h_{r_i,d_t}$ represent the transmitting power and channel gain respectively of the relay node $r_i$ to the destination node $d_t (d_t = d_1, d_2, ..., d_T)$.

The goal of the above process can be expressed as below. For each relay node $r_i$ in the network, choose a channel from a source node to the destination node aimed to maximize the $T_{r_i,d_t}$, and then assign the relay node $r_i$ to the corresponding source and destination nodes. Until all of the relay nodes are distributed, each source node can get the information of the corresponding relay nodes.

*2) Obtain the common free spectrum of each source node $s_t$ (user) and the corresponding relay nodes $R_{s_t}$*

Based on the spectrum sensing results of source nodes and relay nodes in the network, we can get the common free spectrum nodes $N_{s_t}$ for each source node $s_t$ (user) and relay nodes $R_{s_t}(s_t = s_1, s_2, ..., s_T)$, where $N_{s_t} = \{n_1^{S_t}, n_2^{S_t}, ..., n_L^{S_t}\}$, and $L$ is the number of common free spectrum. The spectrum selection process can be shown as below.

```
for n=1:N
  if n only idle for s_t and R_{S_t} then
    N_{S_t} ← N_{S_t} + {n}
  else if n idle for several s_t and R_{S_t} (s_t = s_1, s_2, ..., s_T) then
    ϒ_{n,r_i} = (P_{n,r_i} × h_{n,r_i}) / (Γ N_0 W)
    (r_i) ← arg max{ϒ_{n,r_i}}
    if r_i ismember R_{S_t}
      N_{S_t} ← N_{S_t} + {n}
    end if
  else
    drop n
  end if
end
```

where $\Upsilon_{n,r_i}$ represent the SNR of using the idle spectrum n from the relay node $r_i$ to the destination node $d_t$, and can be defined as

$$\Upsilon_{n,r_i} = \frac{P_{n,r_i} \times h_{n,r_i}}{\Gamma N_0 W} \quad (13)$$

*3) allocate spectrum for the purpose of maximizing the network throughput.*

We can get the spectrum prediction results of every idle spectrum $n_l^{S_t}$ in the network,

$$I_{S_t} = \{\{I_{n_1^{S_t}}\}, \{I_{n_2^{S_t}}\}, ..., \{I_{n_L^{S_t}}\}\} \quad (14)$$

where $\{I_{n_l^{S_t}}\}$ is the spectrum prediction results of relay nodes for the spectrum $n_l^{S_t}$. And the value is 1 for occupied, and is 0 for free.

If the spectrum is free for one relay, then the SNR can be obtained through the formula (13), and the SNR accepted by the destination node can be expressed as below,

$$\Upsilon_{n,T} = \sum_{n \in N_{S_t}} \max_{r_i \in R_{S_t}} \Upsilon_{n,r_i} \quad (15)$$

The core ideal of this process is that for each element of $\{I_{n_l^{S_t}}\}$, the free spectrum $n_l^{S_t}$ is assigned to the relay with largest $\Upsilon_{n,r_i}$.

*4) Aggregate the idle spectrum that select a same relay, and complete the cooperative transmission.*

After spectrum allocation, the source nodes and relay nodes aggregate the spectrum that select the same relay. And then the source nodes and relay nodes transmit data by using the aggregated spectrum.

The system throughout of the network can be expressed as

$$T_{all} = \sum_{n=1}^{N} b \log_2(1 + \Upsilon_{n,T}) \quad (16)$$

The flow chart of the algorithm is shown in Figure 6.

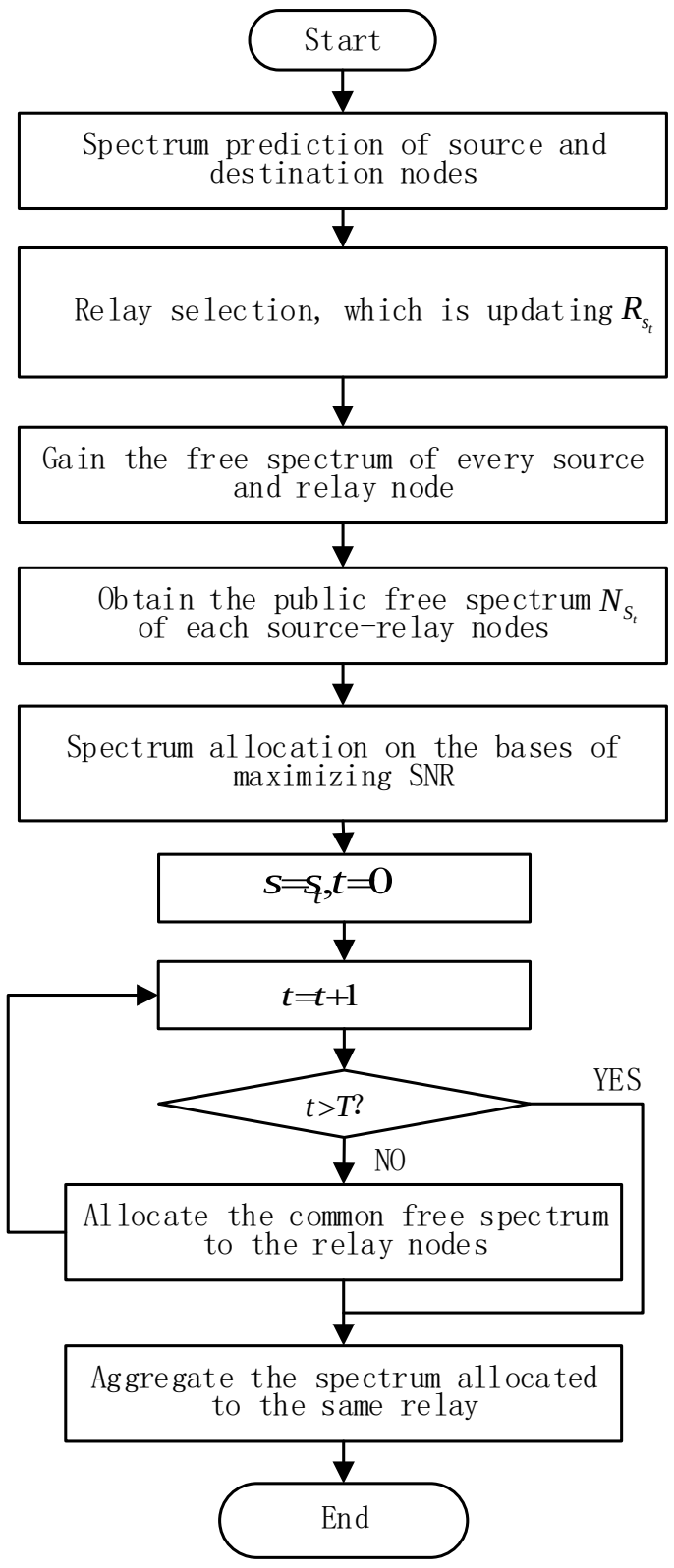

Figure 6. Flowchart of spectrum aggregation scheme in multi-user cooperative relay network

Figure 7. Implementation scheme of spectrum aggregation in cognitive collaboration networks

### *C. Implementation scheme of spectrum aggregation*

In this section, we describe the implementation scheme of spectrum aggregation, and it will cost much to realize the flexible spectrum aggregation schemes [22]. When the available relays are determined, source nodes and relay nodes respectively sense the idle spectrum, and the relay nodes simultaneously send the sensing results to the source nodes. Source nodes collect the sensing results, and combine the spectrum prediction algorithm introduced in last section, then get the public available spectrum and broadcast these information to relays. Relays nodes get the available public spectrum information of themselves, and send RTS signal to destination node. The destination node receive these RTS signal, then estimates the Signal to Noise Ratio (SNR) of each channel and gets the SNR relationship of relay node, spectrum, destination node. The destination node broadcast these information to relay and relay retransmit it destination node. Base on the information of selected available relay set and public spectrum, we select the relay and destination node with the maximal SNR at the spectrum and set assign this relay to the source node to transmit data cooperatively. This process is executed by setting a timer which is in inverse proportion of SNR, for the relay. Assign this spectrum to the relay which is the first one starting the timer and assign remainder spectrum until all the available spectrum is allocated. Through the execution of this spectrum allocation scheme, we can make the system SNR maximal then get the maximal network throughput.

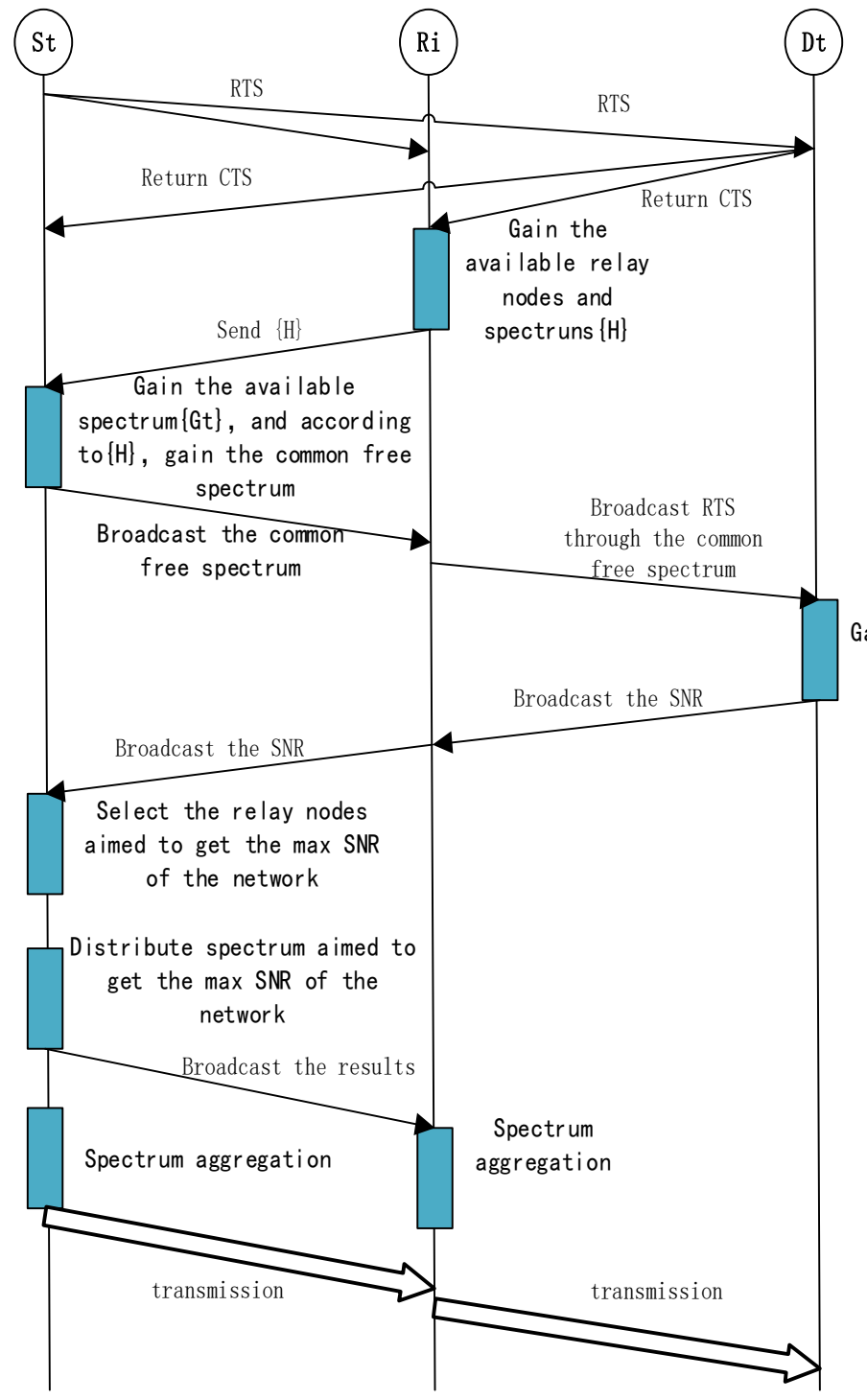

Aiming at maximize the SNR, we get the spectrum allocation scheme. For each relay node, we should estimate if it has been selected by many different available spectrum. If so, we aggregate these available spectrum together, and if not, there will be two cases. If the relay hasn't be selected by any spectrum, this relay will not participate in the communication process this time. If the relay is selected by one available spectrum, this communication process will be executed in this spectrum. When all the spectrum at all the relays has been aggregated, source node transmit data using these spectrum. In the remaining time in the first time slot, source node transmit information to the relay, and at the second time slot, relay retransmit this information to the destination node. A complete communication process is finished. Implementation scheme of spectrum aggregation is illustrated in Figure7.

## IV. SIMULATION RESULT AND DISCUSSION

Based on the analysis above, in this section, we present simulation results for the proposed spectrum aggregation strategy in two steps: first simulate the Markov prediction of spectrum, and then simulate the spectrum aggregation scheme.

### *A. Simulation of Markov Prediction of Spectrum*

We set the simulation process in 100 time slots and execute the simulation process 20 times. For the spectrum state illustrated in Figure5, we simulate the actual spectrum state, default spectrum state, and prediction state in 20 continuous time slots, then we get Comparison diagram of the spectrum state in above three conditions. As shown in Figure 8, at the first 13 time slots the prediction state of the spectrum is the same with the actual state in the network, and the actual state of the spectrum is the same with the default state only at 5 time slots. We can see that using the Markov prediction scheme to predict the state of the spectrum can make the network get more accurate information about the state of the spectrum.

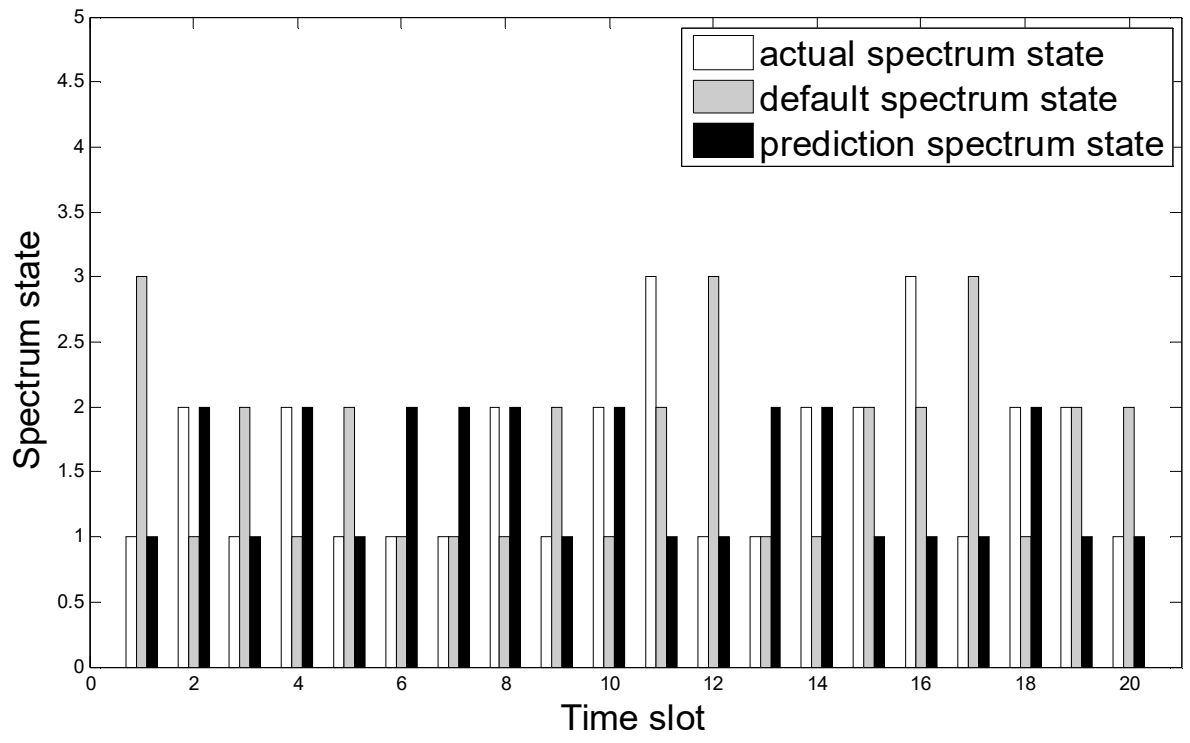

Figure 8. Comparison diagram of actual spectrum state, default spectrum state and predictive spectrum state

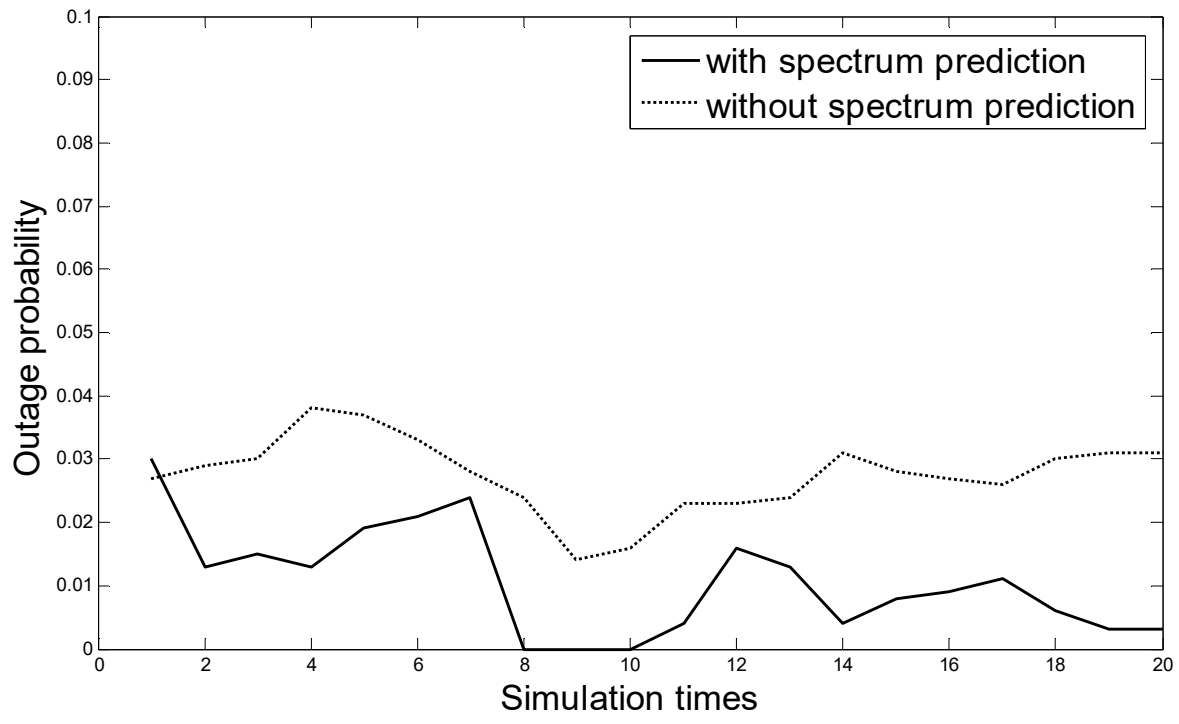

Figure 9. Comparison diagram of outage probability with or without spectrum prediction

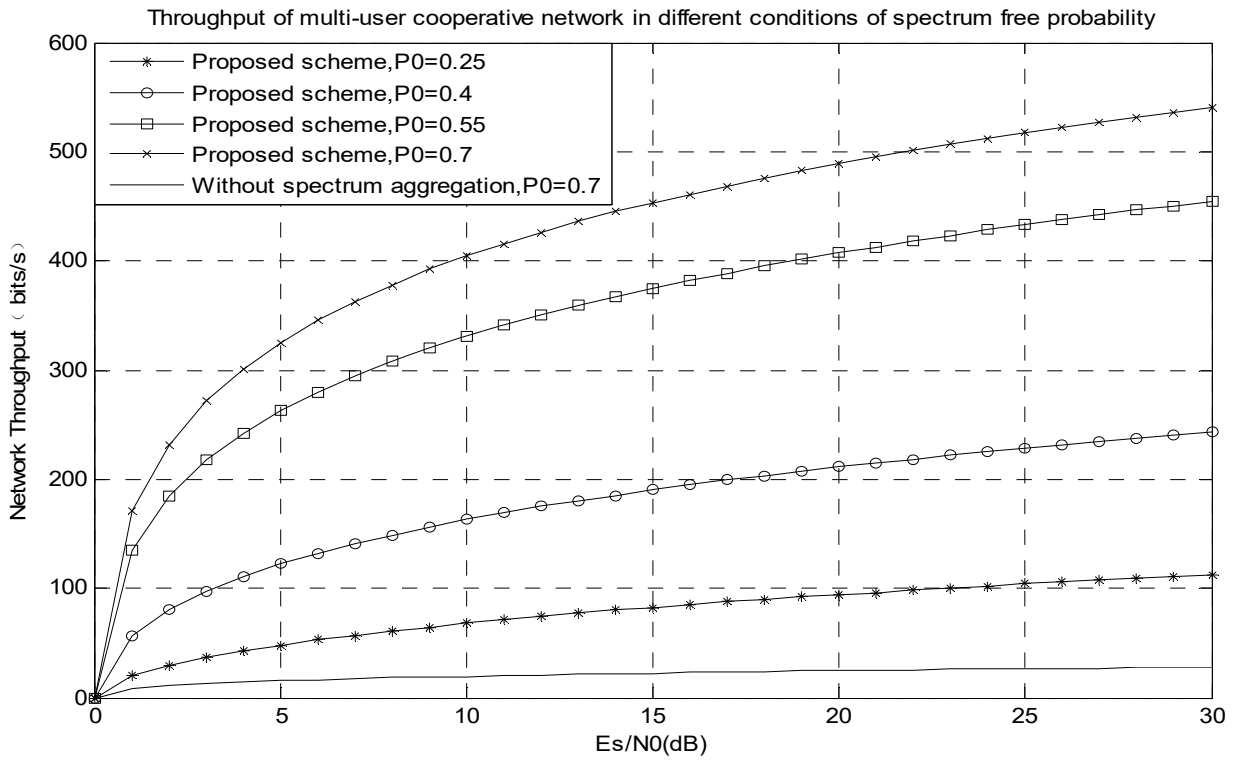

Figure 10. Network throughout with different probabilities of free spectrum

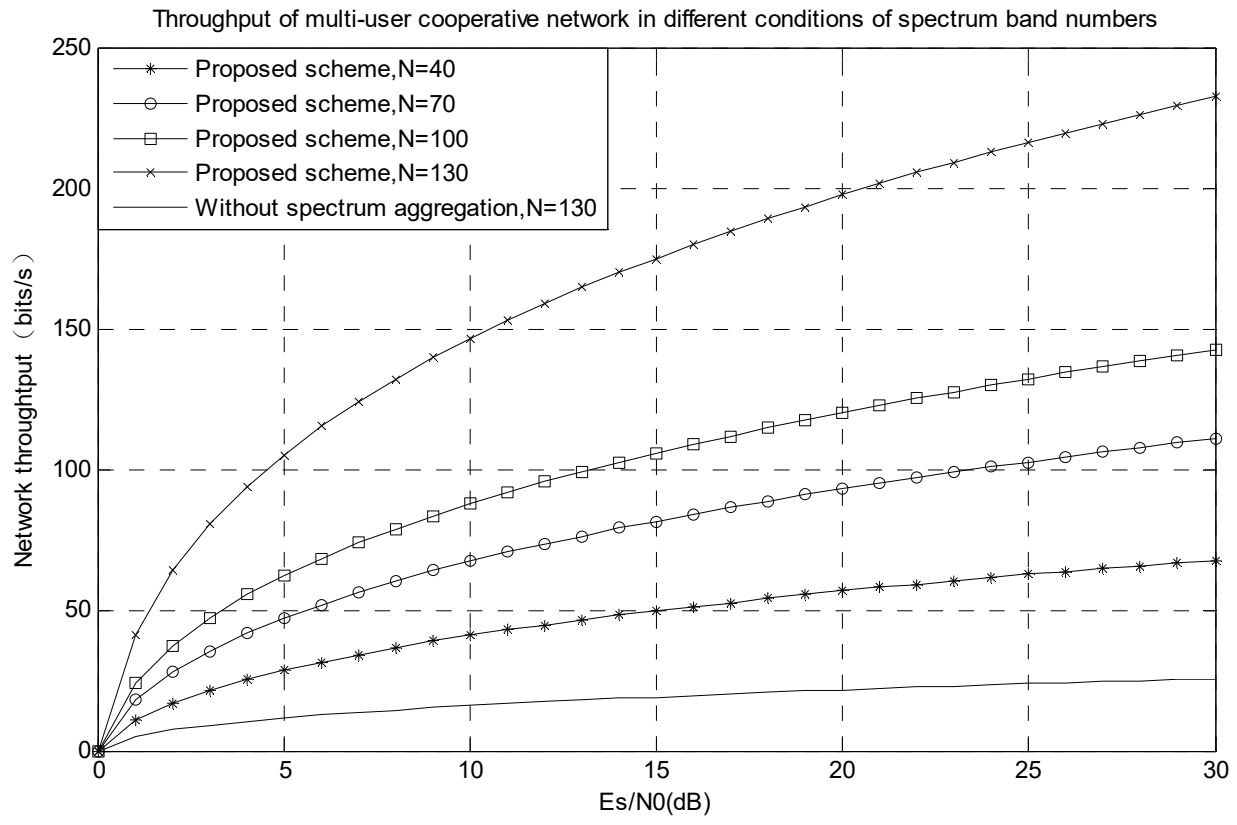

Figure 11. Network throughout with different numbers of spectrum bands

From Figture9 we see that the outage rate in cooperative relay network with Markov spectrum prediction is lower than that without spectrum prediction at the same time slot. It is easy to interpret, when we use Markov prediction model to predict the state of spectrum, for spectrum N, if the sensing state of spectrum at the first time slot is good and the prediction state of the spectrum at the second time slot is good too, but the actual state of the spectrum is busy or bad, it will lead to a high probability of outage of the network. Through the prediction process of the state of spectrum in the cooperative relay networks, due to the accurate prediction results of spectrum, we can choose spectrum with good quality, and the outage rate will be reduced accordingly. At the same time, the spectral efficiency will be improved.

### *B. Simulation of spectrum aggregation algorithm*

The simulation parameters are defined in table 1.In this paper, the simulation of the network throughput mainly considered three variable factors: the number of relay nodes $R_r$, spectrum number N and spectrum idle probability P0. In the figures of the simulation results, the abscissa is the average SNR of straight transmission from the source node to the destination node (if possible), i.e. $E_S / N_0$. Assuming all of the relay nodes are located between the source node and the relay node, and then the $SNR_1$ is the SNR from source node to relay node and the $SNR_2$ is the SNR of relay node to the destination node, they are all less than $E_S / N_0$.

$$SNR_1 + SNR_2 < 2E_S / N_0 \tag{17}$$

In the multi-user cooperative relay network, the simulation mainly compared the throughput of the network using spectrum aggregation algorithm with different parameter conditions, and we also compared to the throughput of the network without using spectrum aggregation as a contrast. The simulation results shows in Figure 5 ~ Figure 7. Through the simulation, the system throughput increases with the increasing of $E_S / N_0$. In the cooperative relay network without using spectrum aggregation, the spectrum idle probability and number of spectrum has little influence on the network throughput, so we only simulate the optimal situation.

As shown in Figure 10, in multi-user cooperative relay networks using spectrum aggregation algorithm, the larger spectrum idle probability is, the higher network throughput will be. This is because when the spectrum idle probability is larger, the available spectrum is more. At the same time, it can be seen in multi-user cooperative relay networks, the throughput of the network using the spectrum aggregation algorithm is higher than that without using the spectrum aggregation algorithm.

As shown in Figure 11, in multi-user cooperative relay networks using spectrum aggregation algorithm, if the number of spectrums is more, then we can get the higher network throughput. Similarly, it can be seen in multi-user cooperative relay networks, the throughput of the network using the spectrum aggregation algorithm is higher than that without using the spectrum aggregation algorithm.

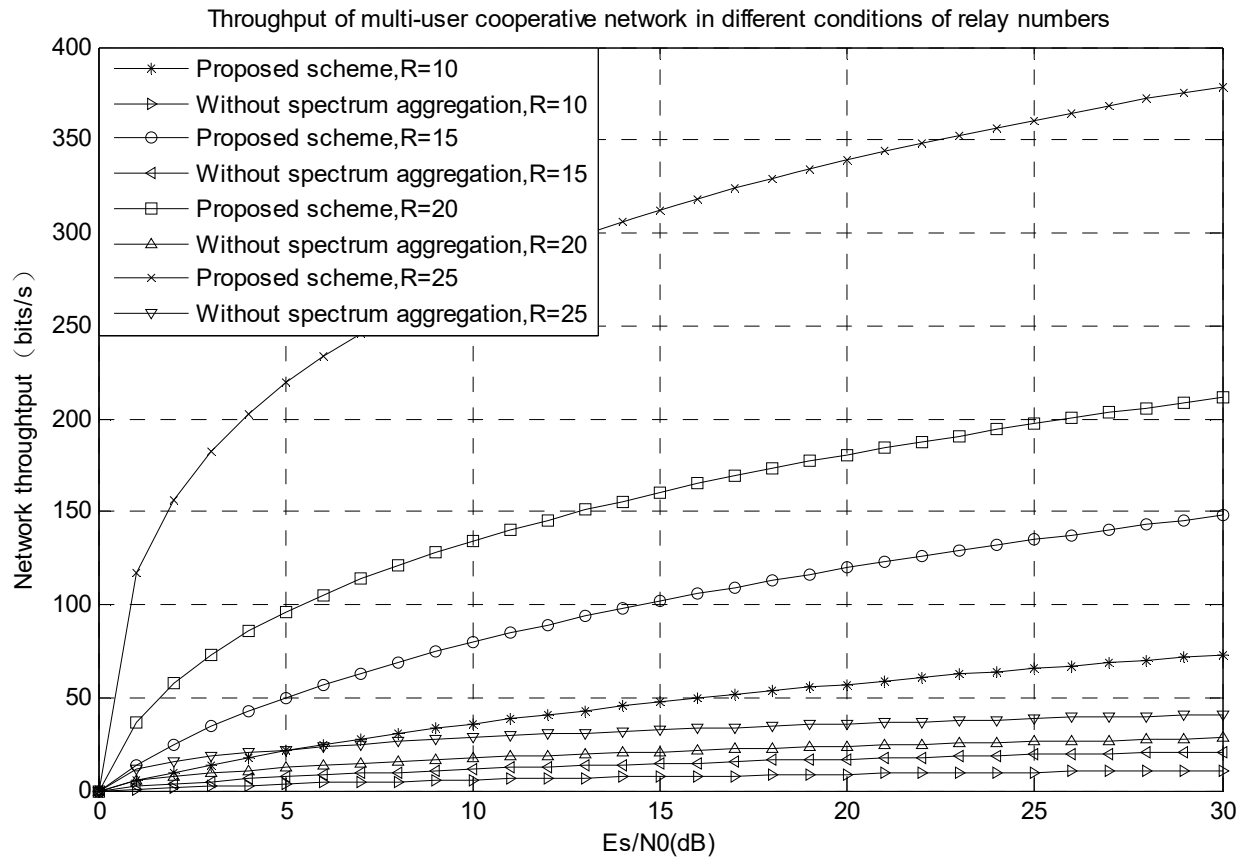

Figure 12. Network throughout with different numbers of relays

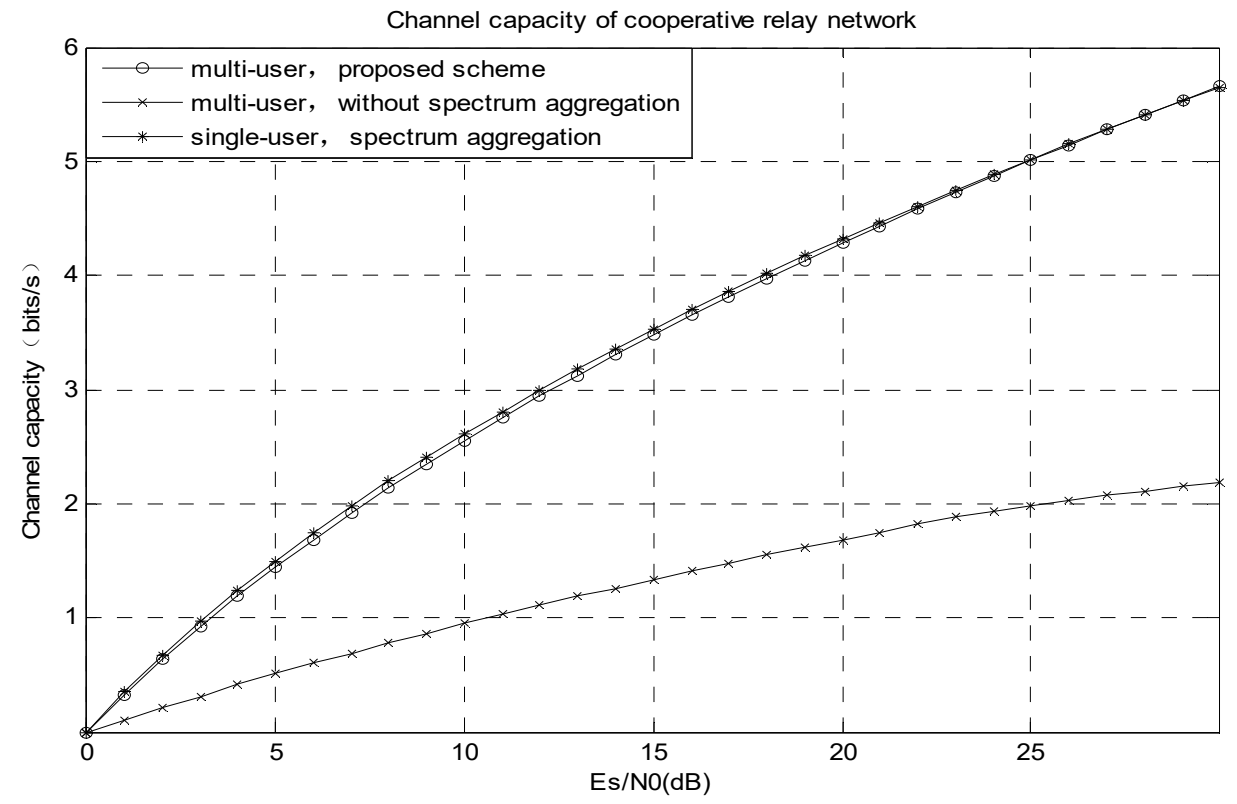

Figure 13. User's channel capacity in the cooperative relay network

TABLE I. SIMULATION PARAMETERS

| Parameters | Values |
|---|---|
| $T$ (numbers of users) | 5 |
| $R_r$ (numbers of relays) | 20 |
| $N$ (numbers of spectrums) | 100 |
| $b$ (spectrum bandwidth) | 2MHz |
| $P_0$ (spectrum idle probability) | 0.4 |

As shown in Figure 12, in multi-user cooperative relay networks, in both cases of whether using the spectrum aggregation algorithm, the network throughput increased with the increasing number of relay nodes. Also, as can be seen from the simulation results, in multi-user cooperative relay networks, the throughput of the network using spectrum aggregation algorithm is much higher than that without using the spectrum aggregation algorithm.

In this paper, we also analyzed the channel capacity of the networks. In a multi-user cooperative relay network with spectrum aggregation, choose a user with the worst performance (that is, the channel capacity is minimal). And in the multi-user cooperative relay networks without spectrum aggregation, select the user of the best performance (that is, the channel capacity is maximal). And compared the above two users' channel capacity with that single-user by using spectrum aggregation. As shown in Figure 13, for the network using spectrum aggregation, the user's channel capacity in multi-user network is not worst than that of single-user network; And in this two situations, the user's channel capacity are both larger than that of the network without using spectrum aggregation.

## V. CONCLUSION

In this paper, based on the Markov Prediction, we propose a spectrum aggregation strategy in the multi-user cooperative relay networks. Taking the spectrum utilization rate and collaboration network into account, we proposed a spectrum aggregation algorithm combined with relay selection. The spectrum prediction process can reduce the network outage rate obviously. The aggregation strategy can improved the spectrum efficiency and transmission rate compared with the traditional cooperative relay networks, through the dynamic relay selection strategy and spectrum aggregation, and also can improve the throughput of the network. Compared with the spectrum aggregation algorithm of single-user networks, the proposed algorithm can improve the channel capacity of the user, and improve the overall performance of the networks.